\documentclass[graybox]{svmult}

\usepackage{type1cm}        % activate if the above 3 fonts are
\usepackage{makeidx}         % allows index generation
\usepackage{graphicx}        % standard LaTeX graphics tool
\usepackage{multicol}        % used for the two-column index
\usepackage[bottom]{footmisc}% places footnotes at page bottom

\usepackage{newtxtext}       % 
\usepackage[varvw]{newtxmath}       % selects Times Roman as basic font

\makeindex             % used for the subject index
\begin{document}

\title{Analysis of the Usability of Automatically Enriched Cultural Heritage Data}
% Use \titlerunning{Short Title} for an abbreviated version of
% your contribution title if the original one is too long
\author{Julien A. Raemy\orcidID{0000-0002-4711-5759} and\\ Robert Sanderson\orcidID{0000-0003-4441-6852}}
% Use \authorrunning{Short Title} for an abbreviated version of
% your contribution title if the original one is too long
\institute{Julien Antoine Raemy \at Digital Humanities Lab, University of Basel, Spalenberg 65, CH-4051 Basel, Switzerland \email{julien.raemy@unibas.ch}
\and Robert Sanderson \at Collections and Scholarly Communications, Yale University, 2 Whitney Ave, New Haven, CT 06510, USA \email{robert.sanderson@yale.edu}}
%
% Use the package "url.sty" to avoid
% problems with special characters
% used in your e-mail or web address
%
\maketitle

\begin{svgraybox}
   This is the preprint version of a chapter submitted to be included in the book "Decoding Cultural Heritage: a critical dissection and taxonomy of human creativity through digital tools", to be published by Springer Nature. The chapter is currently undergoing peer review for potential inclusion in the book.
\end{svgraybox}

\abstract{This chapter presents the potential of interoperability and standardised data publication for cultural heritage resources, with a focus on community-driven approaches and web standards for usability. The Linked Open Usable Data (LOUD) design principles, which rely on JSON-LD as lingua franca, serve as the foundation.
\newline\indent
We begin by exploring the significant advances made by the International Image Interoperability Framework (IIIF) in promoting interoperability for image-based resources. The principles and practices of IIIF have paved the way for Linked Art, which expands the use of linked data by demonstrating how it can easily facilitate the integration and sharing of semantic cultural heritage data across portals and institutions.
\newline\indent
To provide a practical demonstration of the concepts discussed, the chapter highlights the implementation of LUX, the Yale Collections Discovery platform. LUX serves as a compelling case study for the use of linked data at scale, demonstrating the real-world application of automated enrichment in the cultural heritage domain. 
\newline\indent
Rooted in empirical study, the analysis presented in this chapter delves into the broader context of community practices and semantic interoperability. By examining the collaborative efforts and integration of diverse cultural heritage resources, the research sheds light on the potential benefits and challenges associated with LOUD.}

\section{Introduction}
\label{sec:intro}

The success of the International Image Interoperability Framework (IIIF – pronounced “triple-eye-eff”)\footnote{\url{https://iiif.io}}, a model for presenting and annotating digital resources that is backed by a global community developing and maintaining agreed-upon application programming interfaces (APIs) \cite{Snydman-IIIF}, must be learnt from by the cultural heritage sector with respect to the possibility and benefits of widespread interoperability. If, as a community, we can expand from our silos of knowledge into a connected system of interoperable information, the entire sector will benefit, both from the audience perspective of vastly increased access to the information, and also from the publishing perspective of ease of cataloguing and delivery.

This knowledge network would be maintained by GLAM (Galleries, Libraries, Archives, and Museums) organisations as the owners and custodians of cultural and natural history objects, and are thus the best positioned to maintain information about those objects. The publication of that knowledge in an easy to use and consistent methodology will bring about the same ecosystem of tools, usage and understanding as we have seen emerge via IIIF over the last decade. Moreover, IIIF has provided a foundational framework that has not only facilitated the emergence but also guided the development of Linked Art, a community working together to create a shared model based on linked data to semantically describe cultural heritage resources, enabling it to embrace and adhere to analogous structural paradigms, as elaborated in Section \ref{sec:la}. It will facilitate the creation of discovery and research systems without the expense of current aggregators that transform the data, and ensure that the data is kept up to date by incentivising the publishers to do so for their own benefit, rather than for the good of the aggregator. Yale has demonstrated the possibility of this vision through the creation of LUX, described in Section \ref{sec:lux}, which aggregates multiple, independent data sources as Linked Open Usable Data (LOUD), reconciles and enriches the records, and makes a large scale research and discovery system available for unencumbered use.

If more institutions were to publish their data in a consistent and interoperable manner, in order to get the benefits demonstrated by LUX, all institutions’ systems and user experience would be improved by access to the totality of the community's knowledge. This would directly improve exhibitions knowledge with access to all of the host institutions’ and lending institutions’ records, it would allow stronger bibliographic and museum linking such as others’ objects as subjects of published works, and facilitate the creation of digital catalogues raisonnés (such as the Van Gogh World Wide project\footnote{\url{https://vangoghworldwide.org}} and the Duchamp Research Portal\footnote{\url{https://www.duchamparchives.org}}), while better serving critical shared knowledge management tasks such as the maintenance of living artist information. The entire community can contribute their knowledge without significantly changing their existing knowledge management practices, by transforming and using their data according to the LOUD design principles\footnote{\url{https://linked.art/loud}}.

The audiences also directly benefit, be that for teaching and learning, research or general awareness and interest. There are only a few use cases in which the owning institution or physical location is the significant factor, instead the user just wants to discover and interact with the objects via their digital surrogate. If a user is interested in the artist J.M.W. Turner, for example, it is of little concern that “Rain, Steam and Speed” is at the National Gallery in London while “Dort or Dordrecht: The Dort Packet-Boat from Rotterdam Becalmed” is in New Haven, when you can digitally find them via interoperable, semantic descriptions and bring images of them together to compare them side by side where-ever you are through IIIF.

Participating institutions would further benefit via economies of scale. With decentralised data and interfaces, but centralised shared services, such as entity reconciliation and mapping of common datasets, we avoid the challenges of having a single centralised system which inevitably does not perform at scale, costs a lot of money for a single organisation to maintain, and the data is not kept up to date as there is no incentive to do so. However, with centralised shared and standards-based services that could be funded and maintained by the community, we ensure that the functionality is available to all and the costs can be defrayed beyond a single organisation.

That vision might sound like a fanciful fiction at first, but given the impact of IIIF for access to image content, we must consider first why it was so successful, and secondly how we can apply that understanding to advancing broad and usable access to cultural knowledge globally. And for that we must start in the next section with the details of Data Usability.

\section{Data Usability}
\label{sec:data-usability}

Tim Berners-Lee's vision for the Semantic Web \cite{Berners-Lee-SemanticWeb}, or the Web of Knowledge, has been around for almost as long as the Web itself and has been convincingly argued to be ultimately unachievable on a global scale \cite{Target-SemanticWebHistory}, across all knowledge domains. 

Moreover, Bizer et al. \cite{Bizer-LinkedData} identified several persistent challenges for the Semantic Web, a decade after the conception of the Resource Description Framework (RDF), a method for description and exchange of graph data, which include data-driven user interfaces, application architectures, schema mapping, link maintenance, licensing, trust, quality, relevance, as well as privacy concerns. 

However, with the creation of JavaScript Object Notation for Linked Data (JSON-LD) as a developer-friendly serialisation of RDF, we have seen some aspects of that vision realised over the past 10 years. 

\subsection{JSON-LD}
\label{sub:json-ld}

At the time of writing, in September 2023, JSON-LD is used by $45.8\%$ of all websites around the world\footnote{\url{https://w3techs.com/technologies/details/da-jsonld}}. IIIF also uses JSON-LD, although few systems actually depend on the graph that it describes and tend to treat it as JavaScript Object Notation (JSON) of a particular structure. With almost half of all websites using a knowledge graph serialisation, and the success of IIIF in the cultural heritage sector, it is clear that JSON-LD has played a critical role compared to previous attempts.

JSON, as a data syntax and a lightweight data interchange format, is very easy to work with both in the browser and in data management systems. It is compact, and relatively easy to read and scan by the human eye, while enabling nested structures and values that align with programming languages. It can be created by hand in a text editor, or serialised from other data structures using common libraries and tools. This is important because, we argue, the audience for Linked Open Data (LOD) is the developer, and not a researcher or other end user of an application. For LOD to be used, it must be usable, and usability is determined not objectively without context, but instead by the needs and understanding of the user \cite{Sanderson-StandardsCommunities}. The user of the data is the developer, and thus they determine its usability to accomplish their current task, typically to build an application that either publishes or consumes that data to enable discovery and access to the knowledge that it encodes.

\subsection{LOUD Design Principles}
\label{sub:loud-principles}

In his EuropeanaTech 2018 keynote, Sanderson argues that for data to be usable it must have five core features, known as the LOUD design principles and paralleling to some extent Tim-Berners Lee's Five Star Open Data Deployment Scheme\footnote{\url{https://5stardata.info/}}:

\begin{description}[A.]
\item[A.]{It must have the right Abstraction for the audience. The typical approach is to publish in excruciating and incomprehensible detail absolutely everything that is known about a particular topic in a complex structured form. This is neither usable nor necessary for the majority of use cases – the right abstraction of that data is one which allows the user (the developer) to accomplish their task relatively easily and, if at all possibly, enjoyably. If the developer likes to work with the data, then they will continue to do so, and will encourage others to use that format, creating a virtuous cycle. In the same way that the designer of a car's control systems, a mechanic working on it, and the driver all need different access and understanding of that system, so too do different audiences need different access to cultural heritage knowledge.}
\item[B.]{There must be few Barriers to entry. If it is easy to get started, hopefully by merely reading the data and understanding what is happening, then more developers will get started using the data. If it takes a long time to see any sign of progress, many developers will look for an easier route. Conversely, the more people that start and continue to work with the data, the more tools become available, and the more awareness of the data there is. This accelerates the virtuous cycle by demonstrating that not only is it the correct abstraction, it is also quick to accomplish the task.}
\item[C.]{It must be Comprehensible by simply reading the data, rather than having to use specialised tools or require significant initial research to know how to interpret it. A spreadsheet without column headers is incomprehensible, as are formats that rely exclusively on numeric naming conventions for classes, properties or other structures. Uniform Resource Identifiers (URIs) are central in linked data, but URIs should be treated as if they are opaque – users should not read semantics into the URI, and publishers should not feel the need to try and encode details of what the URI identifies within the URI itself. This means that the data must provide some assistance to the user by giving a label or name along every URI.}
\item[D.]{There must be solid Documentation, that has working examples to learn from. While many developers like to get started by reading the data, it is impossible to intuit all of the semantics and possible constructions from looking at examples. There must be solid, easily discoverable reference material that documents very clearly and explicitly what is permissible in the format. That documentation must have examples of each feature, and those examples should be complete and able to be dropped into an implementation of the specification in order to see it in practice.}
\item[E.]{There should be few Exceptions, and instead the data should be internally consistent. Every exception is another rule that the developer needs to understand and then implement in their code. These exceptions are often jarring and uncomfortable to work with, leaving the developer wondering why there is this difference and what other differences there are that they don't yet know about. Conversely, being as consistent as possible means that tools are easy to build and to create testing frameworks to prove that they are correct and complete.}
\end{description}

Overall, the main intention of LOUD is to provide straightforward access to data, primarily for software developers. Thus, a balance must be established that addresses the need for data completeness and accuracy, which depends on the ontological construct, and the pragmatic concerns of scalability and ease of use.

\subsection{Adherence of the IIIF Presentation API 3.0 to the LOUD Design Principles}
\label{sub:iiif-prezi}

The IIIF specifications\footnote{\url{https://iiif.io/api/}} can be easily demonstrated to fulfil all of these requirements for Usability. Taking the IIIF Presentation API version 3.0 as the baseline, its goal is not semantic interoperability, but instead to provide enough information to the audience – the software engineer – to create a view of the object using the referenced images, metadata and other content, i.e. the IIIF Presentation API specifies a standardised description of a collection or compound object (via the \texttt{Manifest} resource) enabling a rich and complex user experience \cite{Appleby-PresentationAPI3}. Comparing this to the above fields, we find that it meets them all easily.

The abstraction of the data is appropriate for the audience to accomplish the expressed task of building a viewing application, as it does not attempt to encode any semantic or descriptive metadata, instead it aligns its structure with that intended usage. Instead of a myriad of metadata fields to understand, it has (label,value) pairs that are divided up by language, in a structure that is easy to read, and easy to code with. It is laid out in such a way that the first part of the data structure is the first part that the developer needs to render to the user, and even URIs are abstracted away into the JSON-LD context document, allowing the developer to deal only with easy-to-read strings and numbers.

\begin{figure}[ht]
\includegraphics[width=1.0\textwidth]{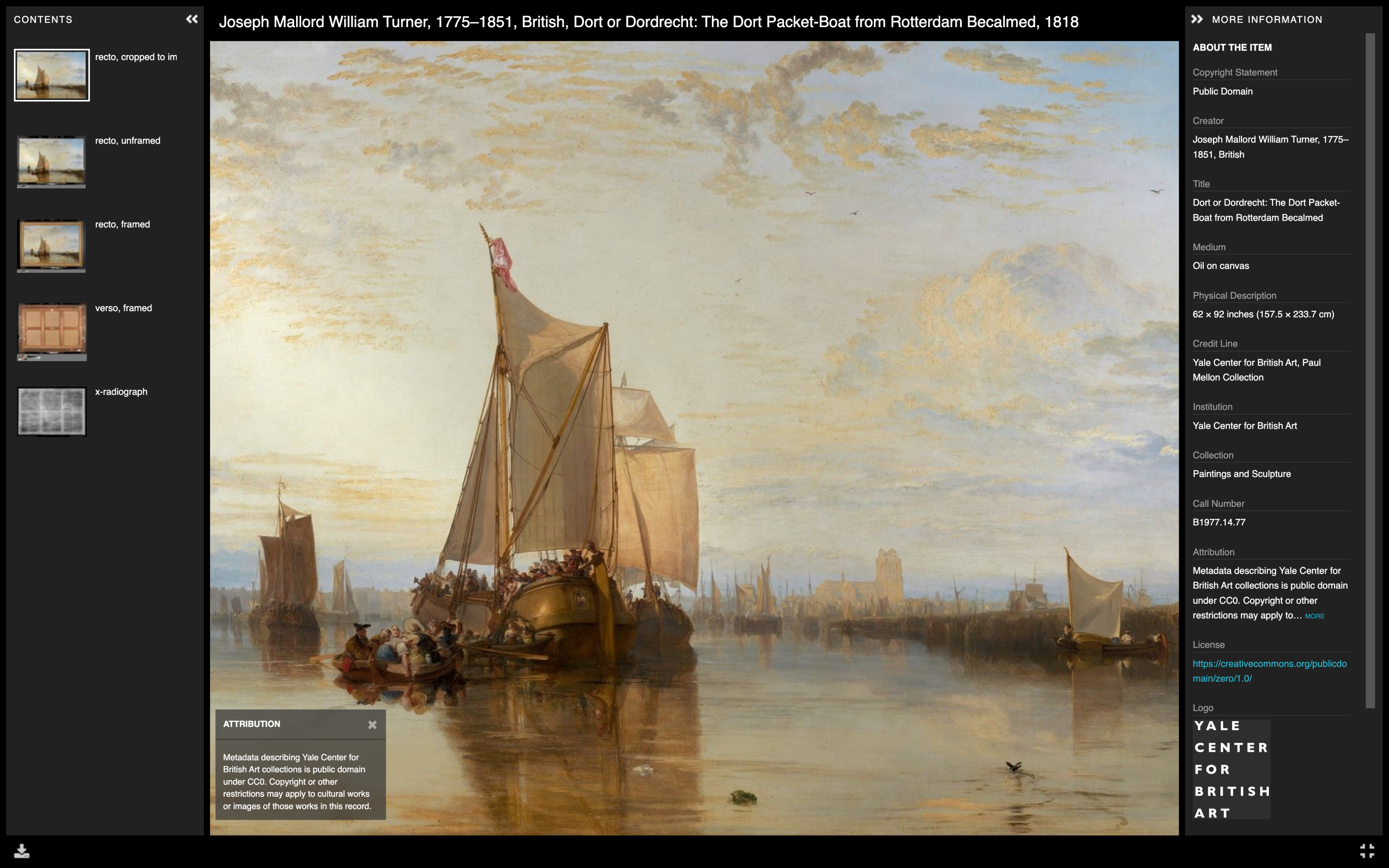}
\caption{\textit{Dort or Dordrecht: The Dort Packet-Boat from Rotterdam Becalmed} by Joseph Mallord William Turner as a IIIF Presentation API 3.0 resource displayed in the Universal Viewer, a IIIF-compliant client. Link to the IIIF Manifest (JSON-LD): \url{https://manifests.collections.yale.edu/v2/ycba/obj/34}.}
\label{fig:dort-uv}       
\end{figure}

There are few barriers to entry to get started with either publishing (a complete instance can be created in just a few lines of code, or easily written by hand), or to build a consuming application. A complete implementation is a lot of work, but to get started is easy, even for someone with minimal programming experience. As it is easy to get started, many have followed through to create wonderful applications that use it.

It is easily understood by reading through the data. The structure and naming conventions are easy to follow, and conform to the expected usage. As JSON, it is a syntax that is very familiar to developers, and that it is also JSON-LD is not even necessary to know, let alone understand, in order to use it\footnote{
Starting in 2018, upcoming IIIF specifications and enhancements to current specifications has embraced JSON-LD 1.1 instead of JSON-LD 1.0. This shift offers numerous advantages, such as the capacity to finely define the impact of context definitions and exert greater control over the specific JSON serialisation. \url{https://iiif.io/api/annex/notes/jsonld/}}.

The documentation is clear and kept up to date. The definition of the structure and possible properties, along with the expected values, are well indexed with expectations for clients and publishers as to what a minimally conforming instance will contain. The examples in the specification itself are not complete, as that would take up a lot of space, however the accompanying cookbook maintains a steady progression of examples and explanations from simple through to the most complex.

Finally, there are few exceptions, even to the way in which images, text, video and other content such as tags or commentary are brought together, in this case through the W3C Web Annotation Data Model \cite{Sanderson-WebAnnotationDataModel}.  The naming conventions of properties, the usage of those properties and the expected usage of them are all clearly defined and do not have special rules based on the context where they are used.

As such, the IIIF Presentation API is, according to the five criteria set out, highly usable and we argue this is the fundamental reason for its success.

\subsection{IIIF Design Principles}
\label{sub:iiif-principles}

The IIIF Design Principles\footnote{\url{https://iiif.io/api/annex/notes/design\_principles/}} express the way in which the IIIF community designed the APIs, leading to their usability. It must be noted that the design principles are not expressed in terms of usability, but as more objective constructs or methodologies that can be followed, which then result in usable data \cite{Appleby-DesignPrinciples}. 

The first design principle is to scope the work through shared use cases. This ensures that the goals of the specifications are clear and well understood, such that the specifications will allow the developer to accomplish their aims, if those aims are met by the use cases that are agreed upon. It also ensures that the specifications are focused on practical details, not on theoretical issues, thereby making them easier to understand. This principle directly deals with requirements A and B.

The next three principles focus on being as easy as possible and keeping the barriers to entry minimal. These deal with requirements A, B and E. By avoiding specific technologies as requirements, and selecting simple (and consistent) solutions, the APIs are easy to get started with, appropriate for the audience, and have few exceptions to deal with.

Principles 2.5, 2.6 and 2.7, 2.9 and 2.10 are about implementation details with the web, and in particular to follow the linked data principles and good web practices. These principles do not fall into the categories above directly, but instead are to ensure that the implementations are performant and fit within existing technologies and standards. 

The closest principle to the notion of usability is perhaps 2.8, which asserts that the specifications should be designed with JSON-LD in mind. The document says that the intent here is to ensure that the data “is as easy to use as possible without the need for a full RDF development suite” and that this will increase the likelihood of adoption. The details of designing for JSON-LD in the IIIF context are then well described later in that document.

The final three principles return to ease of adoption and implementation by ensuring that the different APIs do not all need to be implemented together but instead are loosely coupled, by ensuring that it is internationalised and usable around the world, and that it is easy to extend the specifications to local use cases by defining what is expected to work in which conditions, and leaving everything else unsaid.

\subsection{The Success and Adoption of IIIF}

A number of factors have contributed to the success and uptake of the IIIF. 

First, it hinges significantly on the presence of robust and well-designed software implementations, encompassing both servers and clients. 

Servers are essential for hosting, managing, and serving resources in a manner compliant with IIIF specifications. These server-side software solutions should efficiently handle image requests, metadata retrieval, and other IIIF API interactions while maintaining high performance and scalability. 

Equally important are well-designed, attractive and usable client implementations that are open source and easy to set up as they form the interface through which end-users access and interact with digital resources. For instance, the existence of Mirador\footnote{\url{https://projectmirador.org/}} and the Universal Viewer\footnote{\url{https://universalviewer.io/}} along with the OpenSeadragon\footnote{\url{https://openseadragon.github.io/}} library which they use for dealing with zoomable images, made interoperability an easy case to make to decision-makers and funders. 

As detailed in a study by Raemy in 2023 \cite{Raemy-IIIF-LA}, the IIIF community's success is underpinned by its inclusive and collaborative nature, the availability of interoperable APIs, and compatible implementations. Raemy emphasises the community's openness, friendliness, and commitment to aiding others in their endeavours. Furthermore, the study highlights the collaborative essence of the IIIF community, its connections with prominent figures in the field, and the active participation of technical experts. Raemy also commends the community's well-structured organisation, seamless coordination, and the invaluable support provided by IIIF staff to facilitate cooperation among members. Comprehensive documentation, a pragmatic approach, and the ability to address specific shared needs further contribute to the community's success. The IIIF community's dedication to developing specifications, providing practical solutions and continually evolving the standard underpins its continued appeal.

In light of these attributes that have propelled the IIIF community to prominence, it is noteworthy to delve into a specific aspect of its approach. In striving for widespread adoption of its specifications, the IIIF community undertakes several proactive initiatives, such as writing “cookbook recipes”\footnote{\url{https://iiif.io/api/cookbook/}} to encourage publishers to adopt common patterns in modelling classes of complex objects, enable client software developers to support these patterns, for consistency of user experience as well as to demonstrate the applicability of IIIF to a broad range of use cases. 

Additionally, the community remains highly active, furthering its reach and influence, notably through its various committees and interest groups\footnote{\url{https://iiif.io/community/}}. By consistently seeking advancements and adaptations, the IIIF community not only ensures its relevance but also propels the field forward. This commitment is epitomised by its active exploration of avenues to formally disseminate 3D objects within its framework \cite{Haynes-3D}.

By adopting these easy-to-implement specifications, institutions immediately experience the advantage of not needing to tackle the more complex user-facing components. When considering Linked Art and semantic cultural heritage data, we will look at Yale's LUX from this perspective: if the specifications are easy to publish, is there an adoptable consuming application that demonstrates the value of publishing the data?

While the IIIF Presentation API 3.0 focuses on providing a structured framework with sufficient metadata to facilitate a seamless remote viewing experience, it still doesn't convey semantic information that Linked Art can provide. This highlights a gap that Linked Art bridges by enriching the understanding and integration of cultural heritage data in the digital realm.

\section{Linked Art}
\label{sec:la}

Linked Art is a community-driven initiative collaborating to define a metadata application profile, the model, to describe cultural heritage, and the technical means, a RESTful API, for conveniently interacting with it.  More specifically, it is an RDF application profile of the CIDOC Conceptual Reference Model (CIDOC-CRM) serialised in JSON-LD that incorporates Getty vocabularies\footnote{\url{https://www.getty.edu/research/tools/vocabularies/}}, such as the Arts \& Architecture Thesaurus (AAT), the Thesaurus of Geographic Names (TGN), and the Union List of Artist Names (ULAN), and leverages other commonly used RDF ontologies like RDF Schema (RDFS) and Dublin Core for disambiguating closely related property names used by CIDOC-CRM \cite{Newbury-LOUD}. Linked Art recognises another important perspective: that of software developers who, in many cases in collaboration with scholars, build applications that make use of collections data held by cultural heritage institutions and embrace the LOUD design principles \cite{Page-LinkedArt}.

The goal of Linked Art is to use linked data to enhance cultural heritage collections, particularly focusing on artworks and their origins. This approach enables consistent and structured ways for art institutions to share art-related data. Since it is based on the high-level ontology CIDOC-CRM, which is developed and maintained by the International Committee for Documentation of the International Council of Museums \cite{Doerr-CIDOC}, Linked Art describes assertions in an event-centric paradigm rather than a conventional object-centric framework. Thus, any activity can be potentially represented in an event-centric ontology and is advantageous for modelling temporal data, enabling better discovery of relationships as well as facilitating fine-grained tracking of changes and historical analysis.

\begin{figure}[b]
\includegraphics[width=1.0\textwidth]{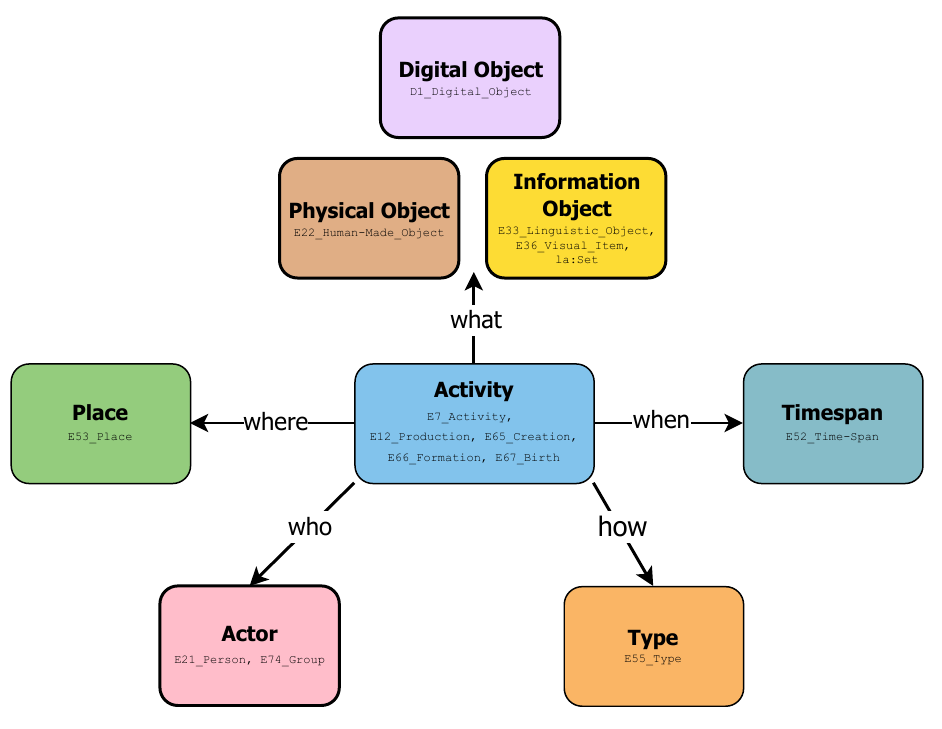}
\caption{Linked Art from 50,000 feet \cite{Raemy-LinkedArtWorkflow,Sanderson-ImportanceOfLOUD}.}
\label{fig:la-50k}       
\end{figure}

\subsection{Conceptual Model}

Linked Art\footnote{\url{https://linked.art}} is documented, much like IIIF, in an incremental approach where common use cases of stakeholders – compiled through GitHub issues in a trasparent manner\footnote{\url{https://github.com/linked-art/linked.art/issues}} – greatly influence the model \cite{Raemy-LinkedOpenUsableData}. 

Figure \ref{fig:la-50k} shows the high-level conceptual model of Linked Art. It comprises some of the CIDOC-CRM classes leveraged by Linked Art. The model primarily addresses five of these provenance questions: “what”, “where”, “who”, “how”, and “when”, akin to some extent to the W7 model developed by Ram and Liu to capture provenance semantics \cite{Ram-SemanticsDataProvenance}.

The model consists of various interconnected components, some of which share common patterns, while others have unique patterns tailored to their specific characteristics. When working with an open ontology like CIDOC-CRM, having these common baseline patterns is valuable. They have been established through experience with datasets from numerous museums, offering practical ways to structure cultural heritage data.

There are a few core properties that every resource should have for it to be a useful part of the world of linked data:

\begin{description}[@contextt]
    \item[\texttt{@context}]{Contains a reference to the context mapping which determines how to interpret the JSON as LOD. It is not a property of the entity being described, but of the document. It must be present.}
    \item[\texttt{id}]{Captures the URI that identifies the object. Every resource must have exactly one id, and it must be an HTTP URI.}
    \item[\texttt{type}]{Captures the class of the object, or \texttt{rdf:type} in RDF. Every resource must have exactly one class. This allows software to align the data model with an internal, object oriented class based implementation.}
    \item[\texttt{\_label}]{Captures a human readable label as a string, intended for developers or other people reading the data to understand what they are looking at. Every resource should have exactly one label, and must not have more than one. It is just a string, and does not have a language associated with it -- if multiple languages are available for the content, then implementations can choose which is most likely to be valuable for a developer looking at the data.}
\end{description}

Additionally, CIDOC-CRM functions as a framework that needs to be extended through the utilisation of additional vocabularies and ontologies to become useful. The provided mechanism for achieving this is the \texttt{classified\_as} property, which points to a term from a controlled vocabulary. This is in contrast to the \texttt{type} property mentioned earlier, which is reserved for CIDOC-CRM defined classes and a few specific extensions as required.

Below is a JSON-LD snippet example of an assertion stating that this object is a painting and, therefore, an artwork, using AAT terms.

\begin{verbatim}
{
  "@context": "https://linked.art/ns/v1/linked-art.json",
  "id": "https://linked.art/example/object/20",
  "type": "HumanMadeObject",
  "_label": "Simple Example Painting",
  "classified_as": [
    {
      "id": "http://vocab.getty.edu/aat/300033618",
      "type": "Type",
      "_label": "Painting"
    },
    {
      "id": "http://vocab.getty.edu/aat/300133025",
      "type": "Type",
      "_label": "Work of Art"
    }
  ]
}
\end{verbatim}

Further identified patterns within the conceptual model, all vetted by the Linked Art community, consist of object descriptions, people and organisations, places, digital integration (such as leveraging the IIIF specifications), provenance of objects, collections and sets, exhibitions of objects, primary sources of information, assertion level metadata, and dataset level metadata. Each pattern plays a pivotal role in defining and organising data related to artworks, artists, locations, digital assets, historical contexts, collections, exhibitions, and the metadata that underpins this interconnected web of cultural heritage data.

\subsection{API Design Principles and Requirements}

Linked Art also follows the footsteps of IIIF in terms of scoping how web specification should be developed by defining its own sets of API design principles and requirements\footnote{\url{https://linked.art/api/1.0/principles/}}.

The design principles are rooted in practicality and interoperability. They are crafted with shared, well-understood use cases, ensuring that the resulting specifications solve real-world problems. Internationalisation is prioritised to remove language barriers for users. The APIs aim for simplicity, allowing for both basic and complex use cases, with the flexibility to start small and incrementally build up. They avoid dependency on specific technologies, making them adaptable across various implementations. By following REST principles, they seamlessly align with the web, ensuring easy caching and interaction. JSON-LD serves as the primary serialisation method, promoting user-friendly representations. Whenever possible, Linked Art adheres to existing standards and best practices to integrate seamlessly with the broader web-based cultural heritage data landscape. Extensibility is encouraged, enabling experimentation and early adoption of new versions. Lastly, Linked Art embraces the network's role in information access, recognising that a multitude of publishing environments is more valuable than overly simplistic consuming implementations.

The Linked Art API requirements are grouped into four key areas, further illustrate how Linked Art aims to provide implementation-based guidance for creating specifications that are not only practical but also responsive to the needs of the cultural heritage sector.

\begin{description}[Consistency across Representations]
    \item[Trivial to Implement]{Linked Art adheres to principles that prioritise ease of implementation, allowing data to be generated without the need for databases or dynamic systems.} 
    \item[Consistency across Representations]{It is maintained by ensuring that each statement appears in only one response document, if possible. Moreover, If a resource has references from multiple other resources, then it needs to be in its own response. Lastly, an efficient handling of inverse relationships is required as each connection should be encoded in a one-way direction, although Linked Art considers exceptions for performance and easy data access through a separate API for some cases.}
    \item[Division of Information]{It focuses on representing 1-to-many relationships from the “many” side, defining deterministic and straightforward rules for data representation, and embedding resources when they have a 1:1 relationship with their parent to reduce the number of separately maintained resources.} 
    \item[URI Requirements]{This requirement stipulates that resources not requiring separate dereferencing do not need their own URIs, and the flexibility of URI structure is maintained, allowing for a broad range of implementations without specific URI structure requirements for API endpoints. } 
\end{description}

If there aren’t any specific URI structure requirements, there are best practices for URIs documented within the Linked Art protocol\footnote{\url{https://linked.art/api/1.0/protocol/}} with preferred endpoint paths. The top-level entity endpoints\footnote{\url{https://linked.art/api/1.0/endpoint/}} align mostly with the core classes of the Linked Art model. At the time of writing, there are eleven endpoints, loosely based on the conceptual model presented previously: 

\begin{description}[Provenance Activities]
    \item[Concepts]{Types, Materials, Languages, and others, as full records rather than external references}
    \item[Digital Objects]{Images, services and other digital objects}
    \item[Events]{Events and other non-specific activities that are related but not part of other entities}
    \item[Groups]{Groups and Organisations}
    \item[People]{Individuals}
    \item[Physical Objects]{Physical things, including artworks, buildings or other architecture, books, parts of objects, and more}
    \item[Places]{Geographic places}
    \item[Provenance Activities]{The various events that take place during the history of a physical thing}
    \item[Sets]{Sets, including Collections and sets of objects used for exhibitions}
    \item[Textual Works]{Texts worthy of description as distinct entities, such as the content carried by a book or journal article}
    \item[Visual Works]{Image content worthy of description as distinct entities, such as the image shown by a painting or drawing}
\end{description}

\subsection{Adoption of Linked Art}
\label{sub:la-adoption}

Linked Art, being a relatively novel initiative in comparison to IIIF, has faced challenges in achieving the same level of widespread adoption. The lack of awareness and limited availability of tools and services has hindered broader engagement within the Linked Art community \cite{Raemy-IIIF-LA}.

However, a pivotal moment is on the horizon for Linked Art. Yale’s LUX stands out as a pioneering and substantial implementation, symbolising a turning point and serving as a catalyst for change. LUX, recognised as a flagship initiative, effectively showcases the substantial potential and transformative influence embedded in the Linked Art and IIIF specifications.

The valuable insights and advancements brought forth by LUX hold the promise to reshape the prevailing perspectives within the community. In the subsequent section, a detailed exploration into the transformative impact of LUX ensues, shedding light on its potential to shape perceptions, and importantly, its role in potentially fostering increased adoption of portals implementing standards that adhere to the LOUD design principles.

\section{LUX: Yale Collections Discovery}
\label{sec:lux}

LUX\footnote{\url{https://lux.collections.yale.edu/}} is an implementation of Linked Art and IIIF as a discovery and research platform for the combined collections of Yale University. This encompasses the Yale Center for British Art (YCBA), the Yale University Art Gallery (YUAG), the Yale Peabody Museum (YPM) and the Yale University Library (YUL). These collections encompass art, natural history, bibliographic and archival collections, and all of the related people, organisations, places, concepts and events, totalling some 41 million records at the time of writing.

\begin{figure}[ht]
    \centering
\includegraphics[width=1.0\textwidth]{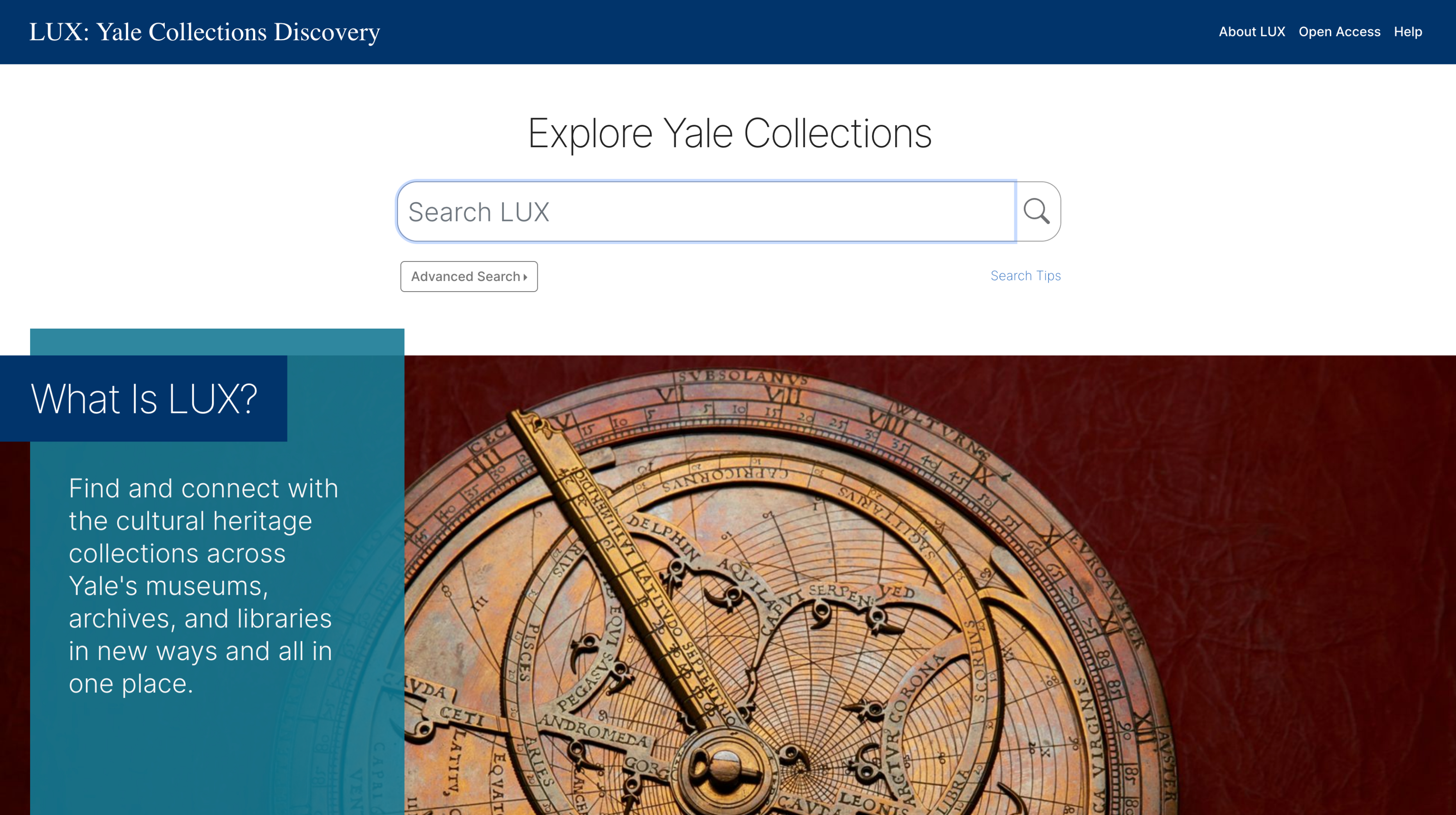}
    \caption{LUX Homepage}
    \label{fig:enter-label}
\end{figure}

Beyond just using the Linked Art metadata application profile, the development of LUX also tried to apply the same principles to other decisions that were needed when mapping data and use cases from the systems of record into Linked Art, and which functionality was important to implement.

The system consists of several interconnected components, namely data harvesting, data pipeline, back-end database, middle tier and front-end. These components have been integrated according to established standards including both IIIF and Linked Art, so that any individual component can be replaced without requiring a complete rewrite of the system.

\subsection{Developing a LOUD-driven discovery platform}
\label{sub:lux-dev}

The usability of the data was extremely important to the development process as it meant that a relatively junior front-end software engineer was able to build the application without significant assistance. The data format being easy to understand and work with meant she could dive in and get started and stakeholders could immediately see results. The consistency of the structure meant that components could be built that leveraged the repeated patterns and then could be reused whenever that pattern was encountered. By following the design principles adopted from IIIF, the implementation architecture meant that the resulting system is performant, scalable, relatively modular and easy to adopt and adapt. 

Discussions around data mapping decisions were easier given the design principles and specifications. Instead of the discussions being about competing viewpoints, which has often led to frustration and lack of engagement, instead they could be structured around cooperatively determining which possibilities best aligned with the principles, and which were outside them. Examples of requested modelling that was determined through this process to be outside of the usability guidelines, and therefore out of the scope of the work, was a desire to align the parallel structures of textual description and structured data around dimensions and materials of an object, the inclusion of meta-meta-data such as the provenance of where individual assertions came from in the merged records, and structured data around uncertainty of assertions. Cases that would have led to inconsistency and more exceptions in the mapping included that animals referenced as subjects or actors could be treated as people, and fossils in the natural history museum could be treated as human made objects. Without the structures to help focus the attention on usability rather than correctness and completeness, these situations all would have led to either long and fraught discussions or aggravated developers needing to deal with more and more complex data structures.

This paradigm also helped with determining the correct approach for systems architecture and functionality.  The hardest challenge of using a knowledge graph was the need to have a traditional records style interface with keyword search, facets and views of the individual entities. A triplestore or native graph based system does not easily enable any of these, and requires multiple systems to be used in conjunction, which increases the complexity of development and maintenance of the platform. Instead after several months of research, a multi-modal platform was licensed which can treat the records as records, and extract the relevant parts of the graph and allow a single query to use the features of both worlds simultaneously.

Again following the principles such as as simple as possible but no simpler, the graph parts of the queries were analysed against the search requirements and the resulting relationships simplified to only what was necessary. As the record maintains the full data, no information is lost, however the performance and ease of development was increased by collapsing complex chains of relationships down to only one artificial predicate. For example, in order to capture the role of each artist in the production of an artwork, the object is produced by a \texttt{Production} event, which then has parts to represent the roles, and each part is carried out by a \texttt{Person} or \texttt{Organisation}. To simplify the common query of objects produced by a given artist, that was reduced to the equivalent of the Dublin Core relationship of creator in the graph. This pattern was then applied across all of the record types and requirements such that only relationships between records were materialised, resulting in a 40\% reduction in the size of the data and a much more performant system.

One of the principles of Linked Art is that each relationship should only be present in the dataset in one record, including inverse relations. For example, if there are two \texttt{Place} records, and one place is part of the other such as a county within a state, then the part of the relationship is only expressed in the county, and the state does not list all of the counties, cities or other localities which are part of it. This direction is intentional to keep the size of each record down, and relatively consistent. However, it leads to the inevitable and obvious question of how do you determine, in this case, the places which are part of the state? As the solution requires looking up many records, the implementation is a search on that property. To avoid technology dependence on a particular query language or search engine, the Linked Art API makes use of the Hypertext Application Language (HAL) specification \cite{Kelly-JSONHypertext} which allows the record to include links to search URIs and give each a name. The front end need only follow the link to receive a paginated list of all of the results, in the same format as for any other set of search results. This layer of indirection both avoids technology dependence and increases the usability for the front-end developer, who no longer needs to understand the data model and query language in order to retrieve the list of child places, but instead is provided a named link in the record to follow, and a standard response with all of the functionality needed to produce different user interfaces for different situations.

Several other practical choices were facilitated through the LOUD principles and practices. In the LOD world, there is a fascination with federated queries – distributing the query among multiple, potentially heterogeneous systems, and then bringing the results back together before presenting them to the user. This paradigm is unreliable as the speed of the search is dependent on the speed of the slowest participating system, and if any system is offline for some reason, then the results will necessarily be incomplete. The alternative is to harvest all of the data from every participating system and combine it in advance into a single infrastructure. The trade-off is between the extent to which the results are out of date  with the source system, and the speed at which searches can be accomplished by end users. Given the relatively infrequent change of the majority of the records, and that the users' information needs can likely still be satisfied by information that is a day out of date, the harvest approach was selected. The records are made available for synchronisation leveraging the IIIF Change Discovery API 1.0 \cite{Appleby-Change}, with the Linked Art data taking the place of the IIIF resources.  This was significantly easier to implement by the participating libraries, archives and museums than every unit maintaining their own query endpoint.

\subsection{Automatically Enriched Cultural Heritage Data}

\begin{figure}[ht]
    \centering
    \includegraphics[width=1\textwidth]{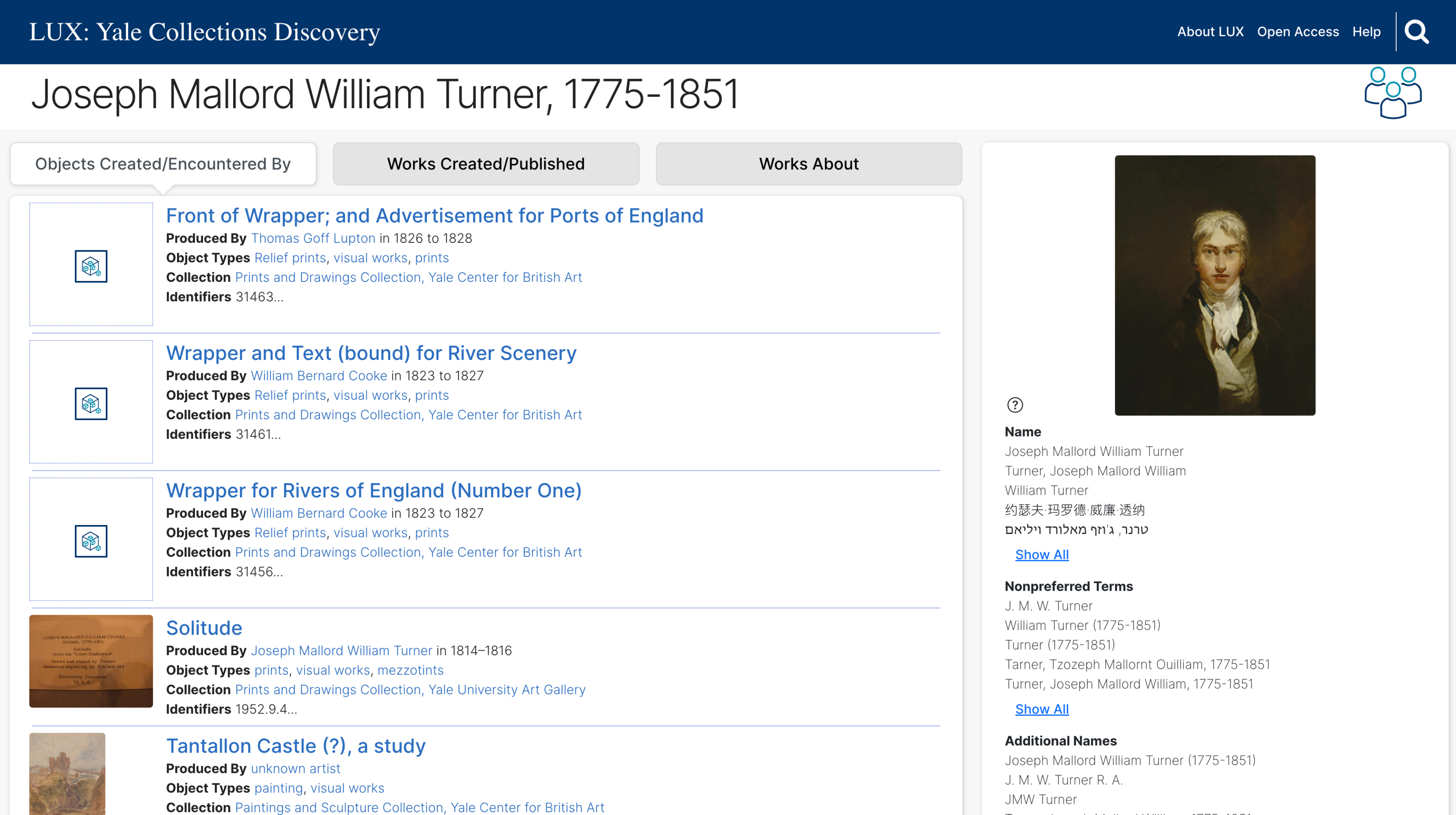}
    \caption{Joseph Mallord William Turner, 1775-1851. The portrait image of Turner is hosted on Wikimedia Commons.}
    \label{fig:turner-person}
\end{figure}

The LUX platform is distinguished by its extensive connections, not only within various Yale units but also outside, as it incorporates external data sources during data processing. These sources encompass a wide range of subject areas and perspectives. This enriches the data accessible to users by matching records within LUX. For instance, one of the key procedures employed to harmonise works and objects involves incorporating reconciled Wikidata records\footnote{\url{https://www.wikidata.org}}, allowing for meaningful connections between items, for instance, in the YCBA collection and related works. Additionally, these sources incorporate additional names and terms from authority records and subject headings, such as those from the French National Library (BnF)\footnote{\url{https://data.bnf.fr}}, the Library of Congress (LoC)\footnote{\url{https://id.loc.gov}}, or the German-speaking Integrated Authority File (GND)\footnote{\url{https://gnd.network}}. In addition, LUX integrates Wikimedia images that are in the public domain, as illustrated by Figure \ref{fig:turner-person}\footnote{\url{https://upload.wikimedia.org/wikipedia/commons/9/9f/Joseph\_Mallord\_William\_Turner\_auto-retrato.jpg}}.

This integration significantly enhances the record by combining knowledge from different Yale units, Getty vocabulary terms, national libraries and other external sources. This is an example of the positive impact and improvement that can be achieved by linking disparate data sources. The linking process is automated within the data processing code, using equivalent URI and intelligent matching of names associated with people, places and things. However, matching and merging data into a single LUX record can be a complex task. Data quality is affected by human imperfections, as all data is derived from human input.

\begin{figure}[ht]
    \centering
\includegraphics[width=0.45\textwidth]{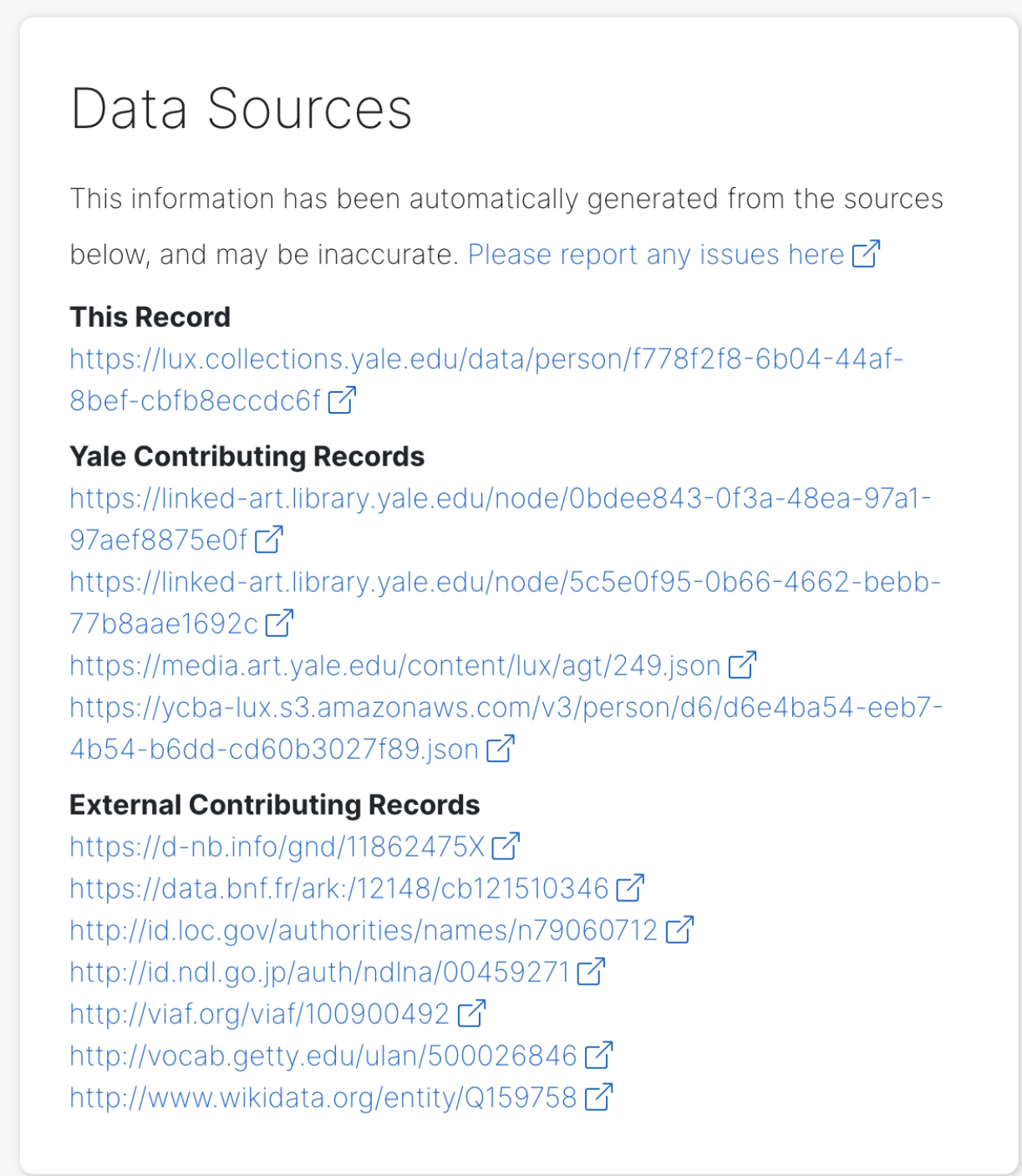}
    \caption{Yale and External Data Sources that have been used for the record about Joseph Mallord William Turner, 1775-1851. Linked Art representation of this record (JSON-LD): \url{https://lux.collections.yale.edu/data/person/f778f2f8-6b04-44af-8bef-cbfb8eccdc6f}.}
    \label{fig:turner-sources}
\end{figure}

\begin{figure}[ht]
\centering
\includegraphics[width=1.0\textwidth]{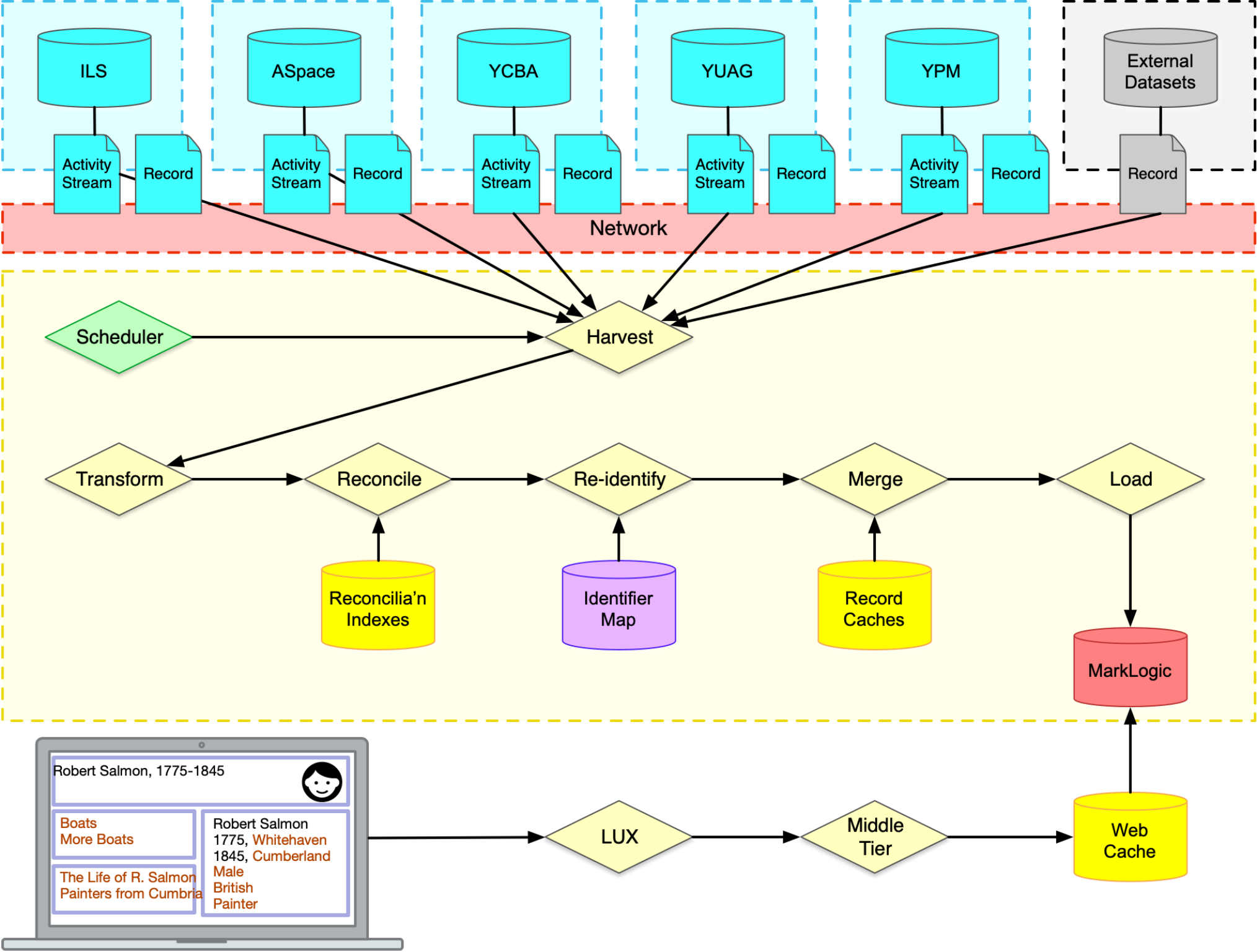}
\caption{LUX Data Pipeline and Architecture}
\label{fig:lux-pipeline}       
\end{figure}

Figure \ref{fig:lux-pipeline} depicts the overall data transformation, reconciliation, enrichment and publication workflow for LUX. The base records come from both internal and external sources, with internal records being harvested via the IIIF Change Discovery API, which is in turn an implementation of the W3C Activity Streams 2.0 specification \cite{Snell-AS}. 

The process (diamonds) named \texttt{Harvest} runs nightly, triggered by an operating system level scheduler to poll each stream to find and retrieve records that have changed since the previous harvest. For external datasets that do not have associated Activity Streams, these records are either retrieved \textit{en masse} via downloadable dump files, or as needed when another record refers to them. The initial state, and all subsequent states after transformations have occurred, are stored in the “Record Caches” store. All records are passed through source specific transformation routines (\texttt{Transform}) in order to either map from arbitrary data formats, or to validate and clean up records already provided in Linked Art. Once the information is available in a consistent format, the records are first sent to a reconciliation engine (\texttt{Reconcile}) to discover further identities from the various datasets to be able to collect all information about a particular entity eventually into a single record. 

Once the records are connected, they have their internal identifiers re-written to a central set of unique identifiers by means of “Identifier Map”, a very fast in-memory database, that maps the original URIs to the internal identifiers (\texttt{Re-Identify}). The result is a transformation of the records where the data remains the same, but the identifiers are now consistent. The records from multiple sources that have been mapped to the same identifier are then merged together (\texttt{Merge}) to form the single record for the entity. The resulting dataset is then annotated with some additional features for indexing and exported to \texttt{Load} into the back-end query engine, a product called MarkLogic\footnote{\url{https://www.marklogic.com/}}, a licensed system by a company called Progress\footnote{\url{https://progress.com/}}.

In order to interact with the data, a user connects to the LUX portal in their web browser and performs a search. That search is sent through a middle tier gateway that allows for seamless transition between MarkLogic installations (a process known as blue/green switching) and through an internal web cache built with Varnish to ensure that repeated queries are only evaluated once. Additional web caches are in place between the user and the LUX front end, including Cloudfront\footnote{\url{https://aws.amazon.com/cloudfront/}}, react cache and the browser's native web cache, to ensure performance is as fast as possible.

After the public launch of LUX in May 2023, the focus for LUX involves developing various services. Among the requested services is direct access to the identifier map of equivalencies, as well as the associated indexes for reconciliation. The open documentation of these services will be of significant value to individuals and institutions within the cultural heritage sector. By providing open access to these resources, LUX facilitates streamlined data reconciliation processes and the creation of meaningful connections between diverse cultural heritage datasets. This accessibility will enable a wider community to make effective use of LUX's capabilities, thereby promoting enriched data and interconnected cultural heritage resources across the sector.

\section{Discussion}
\label{sec:discussion}

This discussion is divided in two parts reflecting on the IIIF and Linked Art communities as well as the development of LUX.

First, we discuss the dimension of community engagement required to create open and interoperable standards, emphasising the collective effort required to create specifications that can be seamlessly integrated into different systems. 

The effectiveness of the second dimension, which focuses on how LOUD standards can facilitate data enrichment, is greatly enhanced by the successful implementation of the first dimension. The collaborative approach to standards development lays the foundation for comprehensive data enrichment processes and highlights the need for standardised approaches to improve data enrichment in different domains.

\subsection{Community Engagement Fosters Open Standards and Interoperability}
\label{sub:community-engagement}

The openness of a standard, while critical, is arguably not sufficient for widespread adoption. Achieving successful interoperability often depends on having a significant platform, either through commercial influence or community involvement as articulated by Nelson and Van de Sompel \cite{Nelson-DLib}:

\begin{quotation}
Because of the growing global adoption of open standards by GLAM institutions, especially IIIF specifications stand as a testimony that rich interoperability for distributed resource collections is effectively achievable. But other promising specifications that aim for the same holy grail are struggling for adoption, and, many times, lack of resources is mentioned as a reason. While that undoubtedly plays a role, it did not stand in the way of rapid adoption of protocols that have emerged from large corporations, such as the Google-dominated schema.org. This consideration re-emphasizes that a core ingredient of a successful interoperability specification, and hence of achieving an interoperable global information web, is a large megaphone, either in the guise of commercial power or active community engagement.
\end{quotation}

For community-driven initiatives like IIIF and Linked Art, they require transparent practices that facilitate on-boarding of new members and governance, such as the establishment of a consortium, to steer the initial vision. However, it is essential to recognise that flexibility is equally paramount, particularly in the early stages of an initiative. Embracing adaptability allows these initiatives to respond to evolving needs and emerging insights from new members, ensuring that the initial vision remains dynamic and responsive to the changing landscape.

Channelling the perspective of the World Wide Web Consortium (W3C) to accomplish a demonstration of (interoperable) implementations \cite{Etemad-W3CProcess}, here is how we define interoperability within a LOUD lens:

\begin{svgraybox}
   Interoperability is a state in which two or more tested, independently developed technological systems can interact successfully according to their scope through the implementation of agreed-upon standards.
\end{svgraybox}

This necessitates the development and availability of several compliant tools and software that adhere to these standards, forming an ecosystem of interoperable solutions. In striving for interoperability, it is essential to recognise the formation of sub-communities or satellite groups, often existing within or adjacent to communities. They play a critical role in the creation and maintenance of tools that align with the specified standards. Heavy reliance on a particular tool, if left without dedicated care and maintenance, can pose significant challenges. In such cases, communities need to rally and take collective action to ensure the tool's sustainability and continued functionality. These instances underscore the necessity for a shared commitment to support and collectively manage critical tools, demonstrating the communal responsibility and collaborative ethos that should define community-driven initiatives.

On a different level, the success of collaboration for developing LUX can be attributed to a shared vision that recognises the value of highlighting the connections between diverse collections across different domains. This shared vision has enabled the resources of all participating units to contribute significantly over a number of years through active participation in committees, working groups and unit-level development efforts.

\subsection{LOUD Standards Facilitates Data Enrichment}
\label{sub:loud-enrichment}

Interoperability and openness stand as essential cornerstones in the facilitation of robust data enrichment processes. The principle of interoperability engenders a collaborative environment wherein disparate data sources and formats converge to augment the quality and completeness of data. This convergence is pivotal in data enrichment, allowing for a seamless flow of data across a spectrum of tools and platforms. Concurrently, openness advocates for unhindered accessibility and availability of data, often epitomised by adherence to open standards. Through the lens of data enrichment, these principles synergistically operate to accommodate a diverse array of sources and perspectives, cultivating a more comprehensive and accurate enrichment process.

The LOUD paradigm embodies the harmonisation of interoperability and openness within data enrichment efforts. By harmonising data enrichment with specifications compliant with the LOUD design principles, data connections are strengthened, ultimately improving both scholarly understanding and user experience by providing enriched, accessible, and contextually interwoven data.

Many cultural heritage institutions have yet to embrace open APIs at scale, which hinders data accessibility and interoperability. Factors such as resource constraints, lack of awareness of the benefits, and complexity of implementation contribute to this slow adoption. Yet despite these challenges, LOUD specifications, such as Linked Art, offer a promising opportunity to address these issues and improve interoperability and data sharing in the cultural heritage domain. As such, Yale's LUX serves as an exemplary model of how combining these specifications, namely IIIF and Linked Art APIs, can provide pathways for robust data pipelines, data reconciliation, and subsequent data enrichment, thereby helping the cultural heritage field to progress.

\section{Conclusion}
\label{sec:conclusion}

Both IIIF and Linked Art, supported by their dedicated and collaborative communities, are strongly committed to advancing the accessibility and interoperability of cultural heritage resources and their associated metadata. These communities actively contribute to the development and maintenance of shared APIs that are critical to promoting the seamless discovery and use of cultural heritage. 

LUX serves as a compelling case study for the use of linked data at scale, demonstrating the real-world application of automated enrichment in the cultural heritage sector. By leveraging linked data technologies, LUX enables users to seamlessly access and explore vast collections of cultural heritage data across Yale's museums, libraries and archives in a single environment. The platform demonstrates how automatically enriched data can improve accessibility, usability and interoperability, ultimately transforming the way users engage with and discover cultural heritage resources.

Achieving semantic interoperability requires the establishment of sound LOUD-compliant ecosystems and workflows \cite{Manz-Workflow}. More specifically, the use of standards such as Linked Art is essential to enable effective data sharing across different domains. In addition, the use of IIIF APIs plays a key role in the seamless delivery and annotation of image-based resources. Importantly, it is possible for institutions of limited resources and size, not necessarily of the scale of larger institutions such as Yale, to achieve this type of interoperability. Collaboration and engagement with the wider IIIF and Linked Art communities becomes critical, providing vital support and expertise, particularly in the absence of human resources or skills.

This combination of collaborative standards and real-world application underscores the potential and need for initiatives such as IIIF and Linked Art to drive transformative progress in the cultural heritage sector and beyond.

\begin{acknowledgement}
We want to express our deep appreciation to the dedicated contributors within the IIIF and Linked Art communities, who have served as a continual source of inspiration for our work. We also extend our thanks to our colleagues at the University of Basel and Yale University for their unwavering support and expertise.
\end{acknowledgement}
%
%%%%%%%%%%%%%%%%%%%%%%%% referenc.tex %%%%%%%%%%%%%%%%%%%%%%%%%%%%%%
% sample references
% %
% Use this file as a template for your own input.
%
%%%%%%%%%%%%%%%%%%%%%%%% Springer-Verlag %%%%%%%%%%%%%%%%%%%%%%%%%%
%
% BibTeX users please use
% \bibliographystyle{}
% \bibliography{}
%

% \begin{figure}[t]
% \sidecaption[t]
% Use the relevant command for your figure-insertion program
% to insert the figure file.
% For example, with the option graphics use
% \includegraphics[scale=.65]{figure}
%
% If no graphics program available, insert a blank space i.e. use
%\picplace{5cm}{2cm} % Give the correct figure height and width in cm
%

% \caption{}
% \label{fig:2}       % Give a unique label
% \end{figure}

\end{document}